\documentclass[dvips]{article}
\usepackage{icrctc07}

\title{H.E.S.S. observations of galaxy clusters}
\shorttitle{H.E.S.S. observations of galaxy clusters}
\authors{W. Domainko$^{1}$, W. Benbow$^{1}$, J. A. Hinton$^{2}$, O. Martineau-Huynh$^{3}$, M. de Naurois$^{3}$, D. Nedbal$^{4}$, G. Pedaletti$^{5}$, G. Rowell$^{6}$ for the H.E.S.S. collaboration.}
\shortauthors{W. Domainko et al}
\afiliations{$^1$Max-Planck-Institut f\"ur Kernphysik, Heidelberg, Germany\\ $^2$School of Physics \& Astronomy University of Leeds, UK\\ $^3$Laboratoire de Physique Nucl\'eaire et de Hautes Energies, Universit\'es Paris VI \& VII, France\\ $^4$Institute of Particle and Nuclear Physics, Charles University, Prague, Czech Republic\\ $^5$Landessternwarte, Universit\"at Heidelberg, Germany\\ $^6$School of Chemistry \& Physics, University of Adelaide, Australia}
\email{wilfried.domainko@mpi-hd.mpg.de}

\abstract{Clusters of galaxies, the largest gravitationally bound objects in the
universe, are expected to contain a significant population of hadronic
and leptonic cosmic rays. Potential sources for these particles are
merger and accretion shocks, starburst driven galactic winds and radio
galaxies. Furthermore, since galaxy clusters confine cosmic ray protons
up to energies of at least 1 PeV for a time longer than the
Hubble time they act as storehouses and accumulate all the hadronic
particles which are accelerated within them.  Consequently clusters of galaxies are potential sources of VHE ($>$ 100 GeV) gamma rays. Motivated by these considerations, promising galaxy clusters are observed with the H.E.S.S. experiment as part of an ongoing campaign. Here, upper limits for the VHE gamma ray emission for the Abell 496 and Coma cluster systems are reported.}

\begin{document}
\maketitle

\section{Introduction}

Galaxy clusters are the largest non-thermal sources in the universe. Radio \cite{giovannini00}, \cite{feretti04} and hard X-ray \cite{rephaeli02}, \cite{fusco04} observations show the presence of accelerated electrons in these systems. It is understood that hadronic cosmic rays accelerated within the cluster volume will be confined there (with energies of up to 10$^{15}$ eV) for timescales longer than the Hubble time \cite{voelk96}, \cite{berezinsky97}. Hence clusters of galaxies act as storehouses for such particles, and therefore a large component of cosmic rays is expected in these systems.

Several sources of cosmic rays can be found in galaxy clusters. Accretion and merger shocks driven by large-scale structure formation have the ability to accelerate cosmic rays \cite{colafrancesco00}, \cite{loeb00}, \cite{ryu03}. Supernova remnant shocks and galactic winds can also produce high-energy particles \cite{voelk96}. Additionally AGN outbursts can distribute non-thermal particles in the cluster volume \cite{ensslin97}, \cite{aharonian02}, \cite{hinton07}.

Due to the expected large component of non-thermal particles, galaxy clusters are potential sources for gamma-ray emission (see \cite{blasi07} for a recent review). Various processes can lead to the production of gamma-ray radiation in these objects. Inelastic collisions between cosmic ray protons and thermal nuclei from the intra-cluster medium (ICM) will lead to gamma-ray emission through $\pi^0$-decay \cite{dennison80}, \cite{voelk96}. Electrons with sufficiently high energies can up-scatter cosmic microwave background (CMB) photons to the gamma-ray range in inverse Compton processes \cite{atoyan00}, \cite{gabici03}, \cite{gabici04}. 

Despite the arguments for potential gamma-ray emission given above, no galaxy cluster has firmly been established as a source of high-energy and very high-energy electromagnetic radiation \cite{reimer03}, \cite{perkins06}.

\section{The H.E.S.S. experiment}

The H.E.S.S. experiment is an array of imaging atmospheric Cherenkov telescopes located in the Khomas highlands, Namibia \cite{hinton04}. It observes in the VHE gamma-ray regime and has a field of view of $\sim$5$^\circ$. Due to the large field of view it is possible to detect extended sources such as supernova remnants \cite{aharonian04}, \cite{aharonian07}. Galaxy clusters are expected to feature extended VHE gamma-ray emission. The H.E.S.S. experiment is well suited to search for such a signal (see e.g. \cite{aharonian04}). 

\section{Targets}

\subsection{Abell 496}

\begin{figure*}
\begin{center}
\includegraphics [width=0.99\textwidth]{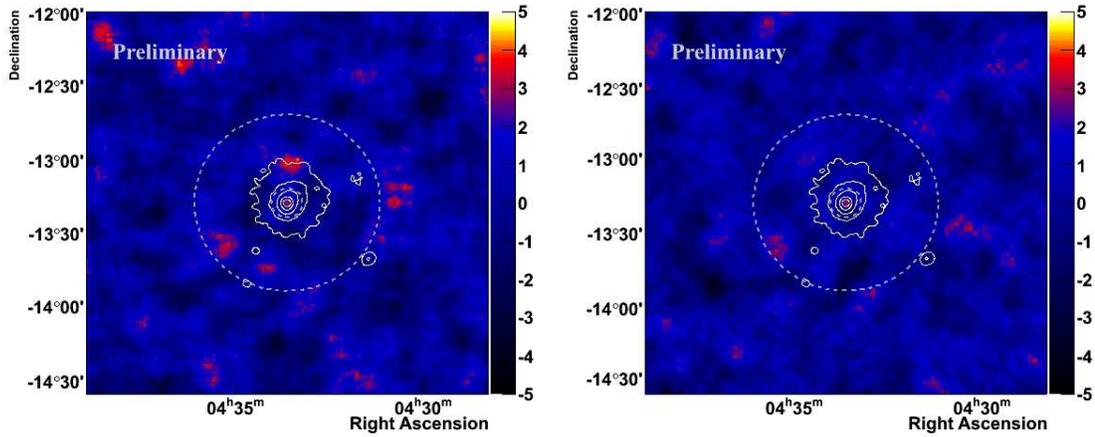}
\end{center}
\caption{Significance map of the cluster Abell 496 seen by H.E.S.S. with standard cuts (left panel) and hard cuts (right panel). No signal is detected from this region of the sky. The dashed circles show the two size cuts and the white contours correspond to \textit{ROSAT} X-ray contours \cite{durret00}.}\label{abell}
\end{figure*}

Abell 496 is a nearby (z = 0.033), relaxed  cluster of galaxies with a mean temperature of 4.7 keV. It features a cooling core at its center \cite{markevitch99}. It is located in the Southern Hemisphere \cite{boehringer04} and is therefore well suited for observations with H.E.S.S. Data taking was performed during moonless nights in the time period from October to December 2005, and in October 2006. In total 23.4 hours of data were taken, with 15.9 hours passing standard data-quality selection (live time 14.6 hours). The mean zenith angle is 27.6$^\circ$ which results in an energy threshold of 0.31 TeV for standard cuts and 0.57 TeV for hard cuts. H.E.S.S. standard data analysis (described in \cite{benbow05}) was performed using different geometrical size cuts to account for the extended nature of the target. No significant excess of VHE gamma-ray emission is found at the position of Abell 496 (see Fig. \ref{abell}). Upper limits for this object for two different size cuts are derived. All upper limits are obtained following \cite{feldman98} assuming a power law spectral index of -2.1 and are given at the 99.9\% confidence level. The first radial size cut, 0.1$^\circ$, is applied to test gamma-ray emission associated with the high density core region of the cluster. This is of particular interest for a hadronic scenario, since the gamma-ray emission should be enhanced in regions with higher density of target material. In this region an upper limit of F$_\mathrm{UL}$($>$0.31 TeV) = $1.0 \times 10^{-12}$ ph cm$^{-2}$ s$^{-1}$ (0.8\% Crab flux) is determined. A radial size cut of 0.6$^\circ$ is also applied, which covers the entire cluster \cite{reiprich02}. For this extended region an upper limit of F$_\mathrm{UL}$($>$0.57 TeV) = $2.4 \times 10^{-12}$ ph cm$^{-2}$ s$^{-1}$ (4.5\% Crab flux) is found. It should be noted that the H.E.S.S. upper limits scale approximately with $r/r_0$ with $r$ and $r_0$ being geometrical size cuts and this relation can be used to convert the presented upper limits to other sizes.

\subsection{Coma cluster}

\begin{figure*}
\begin{center}
\includegraphics [width=0.99\textwidth]{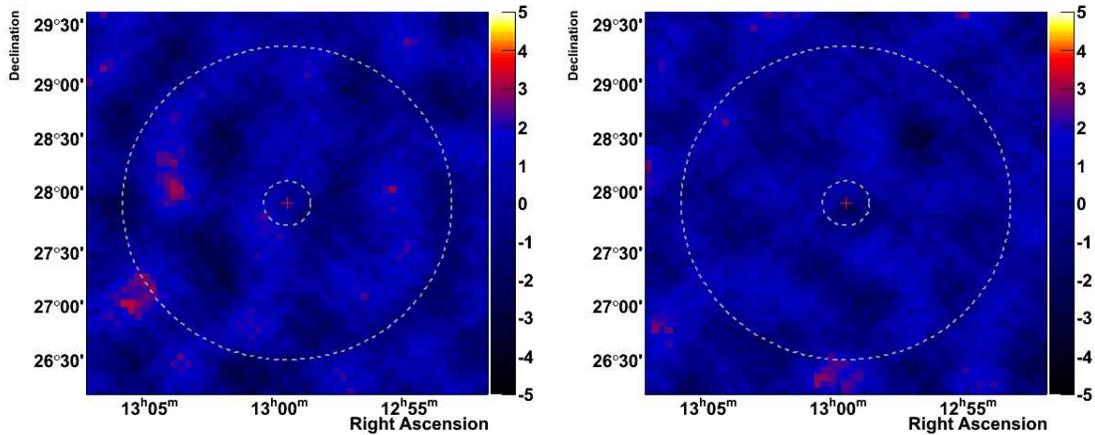}
\end{center}
\caption{Significance map with standard cuts (left) and hard cuts (right) of the Coma cluster. No signal is found in the data. The two dashed circles correspond to the two size cuts.}\label{coma}
\end{figure*}

The Coma cluster is a prominent hot (T = 8.25 keV, \cite{arnaud01}), nearby (z = 0.023) galaxy cluster which shows a merger signature in the X-ray gas (\cite{neumann03}). It features a hard X-ray excess (\cite{rephaeli02}, \cite{fusco04} but see \cite{rossetti04} for a different interpretation)  and a radio halo \cite{giovannini93}. The Coma cluster is {\bf often} considered as a {\bf "standard cluster"}, and, due to the wealth of data on this object, it is very important for theoretical interpretations. It is located in the Northern Hemisphere which makes it less accessible for H.E.S.S. This cluster was observed during moonless nights in April and May 2006. 7.9 hours of good data were obtained resulting in 7.3 hours live time. The mean zenith angle of these observation is 53.5$^\circ$ which results in an energy threshold of 1.0 TeV for standard cuts and 2.0 TeV for hard cuts. No significant signal is found in these observations using various geometrical size cuts (see Fig. \ref{coma}). Upper limits on the VHE gamma ray emission of the Coma cluster for the core region and for the entire cluster are derived. Applying a radial size cut of 0.2$^\circ$ (core region) an upper limit of F$_\mathrm{UL}$($>$1.0 TeV) = $8.3 \times 10^{-13}$ ph cm$^{-2}$ s$^{-1}$ (3.7\% Crab flux) is found. For the entire cluster, with a radial size cut of 1.4$^\circ$, an upper limit of F$_\mathrm
{UL}$($>$2.0 TeV) = $4.8 \times 10^{-12}$ ph cm$^{-2}$ s$^{-1}$ (65.6\% Crab flux) is obtained. For the latter, very extended analysis, only data with a live time of 6.4 hours are used due to an insufficient number of OFF-source runs with such {\bf a} large zenith angle for the background estimation. 

\section{Summary \& outlook}

Clusters of galaxies are the most massive gravitationally bound structures in the universe and as such they are believed to be representatives for the universe as a whole. Therefore they are important tools for cosmology. The detection of gamma-ray emission from these objects will give important information about structure formation and supernova activity over the entire history of the universe. No gamma-ray excess has been found with H.E.S.S. from any cluster, with observation times in the range of 10 $-$ 20 hours. As a next step, one promising galaxy cluster will be given a very deep H.E.S.S. exposure of at least 50 hours.

\section{Acknowledgments}

The support of the Namibian authorities and of the University of Namibia
in facilitating the construction and operation of H.E.S.S. is gratefully
acknowledged, as is the support by the German Ministry for Education and
Research (BMBF), the Max Planck Society, the French Ministry for Research,
the CNRS-IN2P3 and the Astroparticle Interdisciplinary Programme of the
CNRS, the U.K. Science and Technology Facilities Council (STFC),
the IPNP of the Charles University, the Polish Ministry of Science and 
Higher Education, the South African Department of
Science and Technology and National Research Foundation, and by the
University of Namibia. We appreciate the excellent work of the technical
support staff in Berlin, Durham, Hamburg, Heidelberg, Palaiseau, Paris,
Saclay, and in Namibia in the construction and operation of the
equipment.


\begin{thebibliography}{}

\bibitem{aharonian02} Aharonian, F. A. 2002, MNRAS, 332, 215

\bibitem{aharonian04} Aharonian, F. A., Akhperjanian, A. G., Aye, K.-M. et al. 2004, Nature, 432, 75

\bibitem{aharonian07} Aharonian, F., Akhperjanian, A. G., Bazer-Bachi, A. R. et al. 2007, ApJ, 661, 236

\bibitem{arnaud01} Arnaud, M., Aghanim, N., Gastaud, R., et al. 2001, A\&A, 365, L67

\bibitem{atoyan00} Atoyan, A. M. \& V\"olk, H. J. 2000, ApJ, 535, 45

\bibitem{benbow05} Benbow, W. 2005 in Towards a Network of Atmospheric Cherenkov Detectors VII ed. B. Degrange \& G Fontain (Palasieau: Ecole Polythechnique), 163 

\bibitem{berezinsky97} Berezinsky, V. S., Blasi, P. \& Ptuskin, V. S. 1997, ApJ, 487, 529

\bibitem{blasi07} Blasi, P., Gabici, S. \& Brunetti, G. 2007, Int.J.Mod.Phys. A22, 681

\bibitem{boehringer04} B\"ohringer, H., Schuecker, P., Guzzo, L., et al. 2004, A\&A, 425, 367

\bibitem{colafrancesco00} Colafrancesco, S. \& Blasi, P. 1998, APh, 9, 227

\bibitem{dennison80} Dennison, B. 1980, ApJ, 239, L93

\bibitem{durret00} Durret, F., Adami, C., Gerbal, D. \& Pislar, V. 2000, A\&A, 356, 815

\bibitem{ensslin97} En\ss lin, T. A., Biermann, P. L., Kronberg, P. P. \& Wu, X.-P. 1997, ApJ, 477, 560

\bibitem{feldman98} Feldman, G. J. \& Cousins, R. D. 1998, Phys. Rev. D, 57, 3873

\bibitem{feretti04} Feretti, L., Brunetti, G., Giovannini, G., et al. 2004, JKAS, 37, 315

\bibitem{fusco04} Fusco-Femiano, R. Orlandini, M., Brunetti, G., et al. 2004, ApJ, 602, 73

\bibitem{gabici03} Gabici, S. \& Blasi, P. 2003, APh, 19, 679

\bibitem{gabici04} Gabici, S. \& Blasi, P. 2004, APh, 20, 579

\bibitem{giovannini93} Giovannini, G., Feretti, L., Venturi, T., Kim, K.-T. \& Kronberg, P. P. 1993, ApJ, 406, 399

\bibitem{giovannini00} Giovannini, G. \& Feretti, L. 2000, NewA, 5, 335

\bibitem{hinton04} Hinton, J.\,A., 2004, New Astron. Rev., 48, 331

\bibitem{hinton07} Hinton, J. A., Domainko, W. \& Pope, E. C. D. 2007, MNRAS submitted, astro-ph/0701033

\bibitem{loeb00} Loeb, A. \& Waxman, E. 2000, Nature, 405, 156

\bibitem{markevitch99} Markevitch, M., Vikhlinin, A., Forman, W. R. \& Sarazin, C. L. 1999, ApJ, 527, 545

\bibitem{neumann03} Neumann, D. M., Lumb, D. H., Pratt, G. W. \& Briel, U. G. 2003, A\&A, 400, 811

\bibitem{perkins06} Perkins, J. S., Badran, H. M., Blaylock, G., et al. 2006, ApJ, 644, 148

\bibitem{reimer03} Reimer, O., Pohl, M., Sreekumar, P. \& Mattox, J. R. 2003, ApJ, 588, 155

\bibitem{reiprich02} Reiprich, T. H. \& B\"ohringer, H. 2002, ApJ, 567, 716

\bibitem{rephaeli02} Rephaeli, Y. \& Gruber, D. 2002, ApJ, 597, 587

\bibitem{rossetti04} Rossetti, M. \& Molendi, S. 2004, A\&A, 414, L41

\bibitem{ryu03} Ryu, D., Kang, H., Hallman, E. \& Jones, T. W. 2003,
 ApJ, 593, 599

\bibitem{voelk96} V\"olk, H. J., Aharonian, F. A., Breitschwerdt, D. 1996, SSRv, 75, 279

\end{thebibliography}
\end{document}